# Irreversible thermodynamic analysis and application for molecular heat engines

Umberto Lucia [1], Emin Açıkkalp [2]

[1] Dipartimento Energia "Galileo Ferraris", Politecnico di Torino, Corso Duca degli Abruzzi 24, 10129 Torino, Italy; umberto.lucia@polito.it

[2] Department of Mechanical and Manufacturing Engineering, Engineering Faculty, Bilecik S.E. University, Bilecik 11210, Turkey; eacikkalp@gmail.com

**Abstract**

The aim of this paper is to determine lost works in a molecular engine and compare results with macro (classical) heat engines. Firstly, irreversible thermodynamics are reviewed for macro and molecular cycles. Secondly, irreversible thermodynamics approaches are applied for a quantum heat engine with -1/2 spin system. Finally, lost works are determined for considered system and results show that macro and molecular heat engines obey same limitations. Moreover, a quantum thermodynamic approach is suggested in order to explain the results previously obtained from an atomic viewpoint.

**Introduction**

Applications of thermodynamics range from efficiently evaluation of energy resources, relationships among properties of matter, and even living matter. Energy is a thermodynamic property, and it is always constant in the universe; it doesn't destroy but only changes form a form to another one, like work to heat, work to electricity, etc. [1]. The useful work is obtained by evaluating the variations of the energy, which means that any change in a system is always related to a transition between, at least, two different system states.

Energy changes, conversion one form to another form, and interactions with environment are fully explained by the laws of thermodynamics. The first law of thermodynamics expresses the conservation of energy, while the second law states that entropy continuously increases for the system and its environment [2]. The second law gives us information about the quality of the energy as well as allows us to evaluate the irreversibility of any real system [1]. Scientists and engineers have always been trying to obtain the maximum useful work and to decrease losses or irreversibility to the minimum level related to the specific constraints [3-26].

During the last decades, a continuous interest on micro- and nano-thermodynamic cycles is growing, with developments in nano-technology. Thermodynamic assessments for quantum- and nano-systems, hence quantum thermodynamics, has been focused on [27-76]. In recent years, thermodynamic developments in the field of nanotechnology have raised novel questions about thermodynamics away from the thermodynamic limit. Indeed, quantum heat engines are devices that convert heat into work described by the laws of quantum and statistical thermodynamics. They have been a subject of intense research due to their great practical applications, as, for example:

1. Different thermometry approach with the aim to reduce the dimensions of the probe and in pushing away from the thermalisation timescale to obtain a temperature measurement response in the emergence of small thermometers for nano-scale use;
2. The use in opto-mechanical systems for the realization of nano-scale quantum thermal machines with particular interest for nano-mechanical resonators and quantum opto-mechanical engines, which should convert incoherent thermal energy into coherent mechanical work for power applications in photovoltaic systems;
3. The bacteria conversion of light energy into biofuels;

and many others. Unlike a classical heat engine, in quantum heat engines, the energy exchange between the system and thermal reservoirs occurs in quantized fashion. Therefore the quantum heat engines are modelled as sets having discrete energy levels unlike classical engines. But, one of the open problems in this topic is the link between classical and quantum thermodynamics for the analysis and designing of the quantum heat engines.

In 1803, Lazare Carnot developed a mathematical analysis of the efficiency of pulleys and inclined planes [77] in a general discussion on the conservation of mechanical energy. He highlighted that, in any movement, there always exists a loss of "moment of activity". In 1824, his son Sadi Carnot [3] introduced the concept of the ideal engine. It is a system which operates on a completely reversible cycle without any dissipation. But, this result seems nonsensical because, apparently, this system has no irreversibility and, consequently, it could convert all the absorbed heat into work, without any energy loss. On the contrary, Carnot proved that [3,4]:

1. All ideal engines operating between the same two thermal baths of temperature $T_1$ and $T_2$, with $T_1 > T_2$, has the same efficiency $\eta_C = 1 - T_1/T_2$;
2. Any other engine has an efficiency $\eta$ such that always $\eta < \eta_C$.

Consequently, the efficiency of a reversible Carnot cycle is the upper bound of thermal efficiency for any heat engine working between the same temperature limits [3,4,17,77]. Carnot's general conclusion on heat engines is no more than the proof of the existence of natural limit for the conversion rate of the heat into the mechanical energy [4].

A great number of attempts have been developed to improve the calculation of the efficiency of the real machine [5,17,78-86] because all the thermodynamic processes occur in finite-size devices and in finite-time, in presence of irreversibility. The Carnot's limit is inevitable for any natural system [4], and it has always been experimentally verified.

In this paper, it is aimed to determine irreversibilities in a quantum heat engine. Irreversibilities are classified external, internal and total irreversibilities that is sum of internal and external irreversibilities. A method is presented to describe irreversibilities for a quantum heat engine operating -1/2 spin system and then some numerical result are submitted.

**Irreversibility from a quantum point of view**

In This Section we consider the continuous interaction between atomic electrons and the environment photons. For simplicity, but without any loss of generality, we consider the Hydrogen-like atoms in interaction with the electromagnetic waves present in their environment.

The electromagnetic wave is a flow of photons, which incomes into the atoms, are absorbed by the atomic electrons if the electromagnetic wave frequency is resonant, and outcomes from them. At atomic level, the photons can be absorbed by the electrons of the atoms, and an electronic energy transition occurs between energy levels of two atomic stationary states. Then, the photons are emitted by the excited electrons, when they jump down into the energy level of the original stationary state.

Apparently, there are no changes in the energy of the atom, but only in the electronic transition. But, in reality there exists a change in the kinetic energy of the center of mass of the atom, which is usually negligible in relation to the to the energy change in electronic transition. Moreover, the time of occurrence of the energy variation of the atomic center of mass ($10^{-13}$ s) is greater than the time of electronic transition ($10^{-15}$ s).

Here, we stress that an energy variation of the atomic center of mass exists and it cannot be neglected if we consider a great number of interaction as it happens at macroscopic level [87].

Any atomic stationary state has a well defined energy level, identified by the principal quantum number $n$ [88-95]. An electronic transition between two energy levels can occur following the quantum selection rule $\Delta n = n_f - n_i = \pm 1$ [88-95], where the subscript $f$ means final state and the subscript $i$ means initial state. The atom has an atomic number $Z$ and only one electron in the last orbital. This electron moves in its orbital, for which, following the approach used in spectroscopy [93-96], we can introduced [87]:

1. The apparent atomic radius:

$$r_n = \frac{4\pi\varepsilon_0 \hbar^2}{m_e Z e^2} n^2 \tag{1}$$

2. The energy of the atomic level:

$$E_n = \frac{m_e Z^2 e^4}{32\pi^2 \varepsilon_0^2 \hbar^2} \frac{1}{n^2} \tag{2}$$

3. the Sommerfeld-Wilson rule states that [38-41]:

$$\oint p_e \, dr_n = p_e \, r_n = m_e v_e r = n \hbar \tag{3}$$

where $p_e \, r$ is the angular momentum of the electron, being $r_n$ defined by the relation (1), $n = 1, 2, 3, \ldots$ is the principal quantum number, always integer, and $\hbar$ is the Dirac constant, $p_e = m_e v_e$ is the electronic momentum, where $m_e$ is the mass of the electron and $v_e$ its velocity inside the atom, $e$ is the elementary charge, and $\varepsilon_0$ is the electric permittivity. Considering an Hydrogen-like atom, at initial state, the geometric reference system can be fixed in the center of mass of the nucleus, so that the atom is at rest with null momentum $\mathbf{p}_{atm}$. Its Schrödinger's equation is [88-96]:

$$\left[ -\frac{\hbar^2}{2m_N} \nabla^2_{\mathbf{r}_N} - \frac{\hbar^2}{2m_e} \nabla^2_{\mathbf{r}_e} + V(\mathbf{r}_e - \mathbf{r}_N) \right] \psi(\mathbf{r}_N, \mathbf{r}_e) = E_{tot} \psi(\mathbf{r}_N, \mathbf{r}_e) \tag{4}$$

where $\hbar$ is the Dirac constant, $m_N$ is the mass of the nucleus, $m_e$ is the mass of the electron, $\mathbf{r}_N$ is the nucleus coordinate, $\mathbf{r}_e$ is the electron coordinate, $V(\mathbf{r}_e - \mathbf{r}_N)$ is the electrostatic potential, $E_{tot}$ is the total energy, and $\psi(\mathbf{r}_N, \mathbf{r}_e)$ is the wave function. Now, by using the relative coordinates $\mathbf{r} = \mathbf{r}_N - \mathbf{r}_e$, the coordinates of the center of mass $\mathbf{R} = (m_N \mathbf{r}_N + m_e \mathbf{r}_e)/(m_N + m_e)$, the total mass $M = m_N + m_e$, the reduced mass $\mu = (m_N^{-1} + m_e^{-1})^{-1}$, the momentum of the center of mass $\mathbf{P} = M\dot{\mathbf{R}} = -i\hbar \nabla_\mathbf{R}$, and momentum of the reduced mass particle $\mathbf{p} = \mu \dot{\mathbf{r}} = -i\hbar \nabla_\mathbf{r}$, the equation (4) becomes [88-96]:

$$\left[\left(-\frac{\hbar^2}{2\mu}\nabla_\mathbf{r}^2 + V(\mathbf{r})\right) - \frac{\hbar^2}{2M}\nabla_\mathbf{R}^2\right]\psi(\mathbf{r},\mathbf{R}) = E_{tot}\psi(\mathbf{r},\mathbf{R}) \tag{5}$$

The wave function $\psi(\mathbf{r},\mathbf{R}) = \varphi(\mathbf{r})\vartheta(\mathbf{R})$ is usually introduced to separate the equation (5) in the following two equations:

$$-\frac{\hbar^2}{2M}\nabla_\mathbf{R}^2 \vartheta(\mathbf{R}) = E_{CM}\vartheta(\mathbf{R})$$
$$\left(-\frac{\hbar^2}{2\mu}\nabla_\mathbf{r}^2 + V(\mathbf{r})\right)\varphi(\mathbf{r}) = E_\mu \varphi(\mathbf{r}) \tag{6}$$

where $E_{CM} = \mathbf{P}^2/2M$ is the energy of the free particle *center of mass*, and $E_\mu$ is the energy of the bound particle of *reduced mass*, such that $E_{tot} = E_{CM} + E_\mu$, and $V(\mathbf{r}) = -Ze^2/r$.

Now, we consider the Hydrogen-like atom in interaction with external electromagnetic waves. The electromagnetic radiation is a flux of photons, with [96,97]:

1. The energy $E_\gamma$:

$$E_\gamma = h\nu \tag{7}$$

where $h$ is the Planck's constant ($6.62607 \times 10^{-34}$ J s), and $\nu$ is the frequency of the electromagnetic wave;

2. The momentum $p_\gamma$:

$$p_\gamma = \frac{h\nu}{c} \tag{8}$$

We define the thermodynamic control volume as the sphere with center in the center of the atomic nucleus and radius defined by the relation (1) with $n + 1$ instead of $n$. Consequently, the interaction between the electromagnetic radiation and the Hydrogen-like atom can be analysed as the interaction between the flux of photons with an open system (the atom of principal quantum number $n$), through the border of the control volume defined by the sphere of radius:

$$r = \frac{4\pi\varepsilon_0 \hbar^2}{m_e Z e^2}(n+1)^2 \qquad (9)$$

with center in the center of the atomic nucleus. The atomic electron absorbs the incoming photon when its frequency $\nu$ is the resonant frequency, required by the transition between the initial $E_i$ and final $E_f$ energy levels [88-97], corresponding to the quantized energy:

$$\nu = \frac{E_f - E_i}{h} \qquad (10)$$

where $h$ is the Planck's constant. Emission of the this photon results in the reverse process.

The momentum of the incoming photon is $h\nu \mathbf{u}_c/c$, where $\mathbf{u}_c$ is the versor of propagation of the electromagnetic wave, and $c$ is the velocity of light in vacuum. When an electron absorbs the incoming photon, the atomic momentum becomes [88-97]:

$$\mathbf{p}_{atm} = -\frac{h\nu}{c}\mathbf{u}_c \qquad (11)$$

and the electron undergoes an energy levels transition, from the stationary state of energy $E_i$ to the stationary state of energy $E_f$, which results [87]:

$$E_f = E_i + h\nu - \frac{p_{atm}^2}{2M} = E_i + h\nu - \frac{(h\nu)^2}{2Mc^2} \qquad (12)$$

where $p^2_{atm}/2M$ is the kinetic energy gained by the atom, and $M$ is the mass of the atom. So, we can obtain [87-97]:

$$h\nu = \frac{(E_f - E_i)}{1 - \dfrac{h\nu}{2Mc^2}} \tag{13}$$

In a similar way, for the emission of a photon, we can obtain [87-97]:

$$h\nu = \frac{(E_i - E_f)}{1 + \dfrac{h\nu}{2Mc^2}} \tag{14}$$

As a consequence of the absorption of the photon, the laws of conservation of momentum and energy due to the absorption of the photon, hold to [87-97]:

$$\begin{aligned}
\mathbf{P'} &= M\dot{\mathbf{R}}' \\
\mathbf{p'} &= \mu\dot{\mathbf{r}}' \\
\dot{\mathbf{r}}' &= \dot{\mathbf{r}}'_e - \dot{\mathbf{r}}_N = \frac{\mathbf{p}'_e}{m_e} - \frac{\mathbf{p}_N}{m_N} = \frac{m_N \mathbf{p}'_e - m_e \mathbf{p}_N}{\mu M} \\
\dot{\mathbf{R}}' &= \frac{m_e \dot{\mathbf{r}}'_e + m_N \dot{\mathbf{r}}_N}{M} = \frac{\mathbf{p}'_e + \mathbf{p}_N}{M}
\end{aligned} \tag{15}$$

where $\mathbf{p}_N$ is the momentum of the nucleus and $\mathbf{p}_e$ is the momentum of the electron, $\mu = (m_e^{-1} + m_N^{-1})^{-1}$ and $M = m_e + m_N$. Consequently, the Schrödinger's equation becomes [87]:

$$\begin{aligned}
-\frac{\hbar^2}{2M}\nabla^2_{\mathbf{R}}\vartheta(\mathbf{R'}) &= E_{CM}\vartheta(\mathbf{R'}) \\
\left(-\frac{\hbar^2}{2\mu}\nabla^2_{\mathbf{r'}} + V(\mathbf{r'})\right)\varphi(\mathbf{r'}) &= E_\mu \varphi(\mathbf{r'})
\end{aligned} \tag{16}$$

When the photon is emitted, following the same approach, we can obtain [87]:

$$\begin{aligned}
-\frac{\hbar^2}{2M}\nabla^2_{\mathbf{R}}\vartheta(\mathbf{R'}) &= E_{CM}\vartheta(\mathbf{R'}) \\
\left(-\frac{\hbar^2}{2\mu}\nabla^2_{\mathbf{r}} + V(\mathbf{r})\right)\varphi(\mathbf{r}) &= E_\mu \varphi(\mathbf{r})
\end{aligned} \tag{17}$$

with the wave function given as $\psi(\mathbf{r}, \mathbf{R'}) = \varphi(\mathbf{r})\vartheta(\mathbf{R'})$. We can evaluate the energy footprint of the process as:

$$E_{ftp} = \Delta(h\nu) = \Delta E_{CM} = \langle \psi(\mathbf{r}, \mathbf{R}) | H | \psi(\mathbf{r}, \mathbf{R'}) \rangle = \frac{m_e}{M} h\nu \tag{18}$$

where *H* is the Hamiltonian of the interaction. In this way, we have proven that a microscopic irreversibility exists. Indeed, the relation (18) represents the correction term $h\nu/2Mc^2$ related to the irreversibility occurs during the photon absorption-emission process by the electron of a Hydrogen-like atom. This term can be evaluated considering that the energy of an electronic transition is of the order of $10^{-13}$ J, while the energy, $Mc^2$, related to the mass of an atom *M* is of the order of $10^{-8}$ J. So, the correction term $h\nu/2Mc^2$ is of the order of $10^{-5}$-$10^{-4}$ J, negligible compared to 1 in the denominator, obtaining the well known relation (10) used in atomic spectroscopy. But, we must highlight that for a single atom we may not consider this correction because it is very small in relation to the transition energy, but we stress that this energy correction exists, and it is the energy footprint of the process. These results allow us to explain the Carnot's results. Indeed, as a consequence of the continuous interaction between electromagnetic waves and matter, any system loses energy for microscopic irreversibility and, consequently, any system cannot convert the whole energy absorbed into work. Indeed, our result consists in pointing out that the interaction between a photon and an electron in an atom affects the energy level both of the electron and of the center of mass of the atom. When we consider a macroscopic system, we must consider the global effect of an Avogadro's number of atoms ($\sim 10^{23}$ atoms), and the macroscopic effect of the atomic energy footprint for electromagnetic interaction between atoms/molecules and photons results of the order of $10^{19}$-$10^{20}$ J mol$^{-1}$. This energy is lost by matter for thermal disequilibrium, which causes a continuum electromagnetic interaction [73,75,76]. But, this macroscopic irreversibility is no more than a consequence of the microscopic irreversibility. Now, we must consider the macroscopic effect of this microscopic considerations.

**Irreversible thermodynamics analysis**

Entropy may be called as thermal energy cannot be turned into useful work or sometimes, it is called as disorder in the molecular structure. Entropy is a state function and entropy change for a reversible system is:

$$\Delta S = \int \left(\frac{\delta Q}{T}\right)_{rev} = \Delta S_e + \int_0^\tau \dot{S}_g \, dt$$
$$= \Delta S_e + \int_0^\tau \left(\frac{dS}{dt} - \sum_{i=1}^n \frac{\dot{Q}_i}{T_i} - \sum_{in} G_{in} s_{in} + \sum_{out} G_{out} s_{out}\right) dt \quad (19)$$

where $\Delta S_e$ is the entropy variation that could be obtained through a reversible path on which the system exchanges the same fluxes across its boundaries, $\dot{S}_g$ is the entropy generation rate, i.e. the entropy variation due to irreversibility and it represents how considered system [78,98], $\tau$ is the lifetime of the process under consideration, which can be defined as the range of time in which the process occurs [9, 78, 99], and $Q$ is the heat exchanged, $T$ is the temperature of the thermal source, $s$ is the specific entropy and $G$ is the mass flow.

Entropy generation and obtained work from a system are two thermodynamic quantities related one another, because, reversible work is the maximum work that can be provided any system, while actual systems include irreversibility expressed by the entropy generation, that causes reducing at reversible work. Work interactions of a considered system by using of kinetic energy theorem can be written as [71,100,101]:

$$W_{es} + W_{fe} + W_i = \Delta E_k \quad (20)$$

where $W_{es}$ is the work done by the environment on the system, i.e. the work done by the external forces to the border of the system, $W_{fe}$ is the work lost due to external irreversibility, $E_k$ is the kinetic energy of the system, $W_i$ is the internal work, such that [71]:

$$W_i = W_i^{rev} - W_{fi} \quad (21)$$

where $W_i^{rev}$ reversible internal work and $W_{fi}$ lost work resulted from internal irreversibility. $W_{se}$ is the work done by the system on the environment and it can be described as [71,100,101]:

$$W_{se} = -W_{es} - W_{fe} \tag{22}$$

Consequently, relations obtained from the first law of thermodynamics [100,101]:

$$Q - W_{se} = \Delta U + \Delta E_k$$
$$Q - W_i = \Delta U \tag{23}$$

where $U$ the internal energy of the system. $\Delta E_k$ can be written as following [75]:

$$\int_0^\tau E_k \, dt = \int_0^\tau (Q - W_{se} - \Delta U) \, dt = (Q - W_{se} - \Delta U) \tau \tag{24}$$

According to Annila and Salthe, the Noether approach [102] :

$$2 \int_0^\tau E_k \, dt = nh \quad \text{with} \quad n \geq 1 \tag{25}$$

where $n$ multiplies of quanta $h$ is Planck constant. Using eqs. (24) and (25), one can get:

$$(Q - W_{se} - \Delta U) \tau = n \frac{h}{2} = n \pi \hbar \tag{26}$$

where $\hbar$ is reduced Planck constant. Eq. (26) shows relationship Annila and Salthe results and irreversible thermodynamics. According to Gouy-Stodola theorem, total lost work is [10-14]:

$$W^{irrev} = W_{fi} - W_{fe} = T_0 S_g \tag{27}$$

where $T_0$ is the environmental temperature and $S_g$ is the entropy generation. Manipulating previous equations, one can get:

$$(W_i^{rev} - W_{es}) \tau = n \pi \hbar + T_0 \tau S_g \tag{28}$$

Considering an ideal system (without irreversibilities):

$$W_i^{rev} - W_{es} = \frac{n \pi \hbar}{\tau} \tag{29}$$

If electronic transition in an atom is considered for which the difference of the energy $\Delta$ in a cycle, i.e. the absorption and emission of a photon:

$$\Delta = (W_i^{rev} - W_{se}) + (-W_i^{rev} + W_{se}) - \frac{n \pi \hbar}{\tau} + \frac{n \pi \hbar}{\tau} \tag{30}$$

For an irreversible process the equation (30) becomes:

$$\Delta = \left(W_i^{rev} - W_{es}\right) + \left(-W_i^{rev} + W_{es}\right) + T_0 S_g + T_0 S_g - \frac{n\pi \hbar}{\tau} + \frac{n\pi \hbar}{\tau} = 2T_0 S_g \qquad (31)$$

An atom, its ground state of energy is $E_0$, absorbs a photon of frequency $v$ and it passes state of energy $E_1$. Electronic transition is:

$$E_1 = E_0 + hv \qquad (32)$$

Theoretically the atom can have the reverse transition to the ground state

$$E_0 = E_1 - hv \qquad (33)$$

without any footprint $\Delta$ of the process; indeed, considering the cycle of absorption and emission of the photon the footprint results:

$$\Delta = E_1 - E_0 - E_1 + E_0 + hv - hv = 0 \qquad (34)$$

But, in an atom, a bound electron interacts (electrostatic force) with the atomic nucleus and this interaction must have a consequence in the process considered, as we have highlighted in the previous section. When a photon is absorbed by a bound electron, the following transition occurs:

$$E_1 - E_0 = hv + E_{ka} \qquad (35)$$

with $E_{ka}$ the kinetic energy acquired by the atom as a consequence of the energy and momentum conservation. Consequently, it is necessary a greater quantity of energy, respect to the ideal case, to obtain the same transition between the two stationary states. Then, when the photon is emitted the following transition occurs:

$$E_0 - E_1 = -hv + E'_{ka} \qquad (36)$$

with $E'_{ka}$ the kinetic energy in the final state. The footprint of the process can be evaluated as:

$$\Delta = E_1 - E_0 - E_1 + E_0 + hv - hv + E_{ka} + E'_{ka} = E_{ka} + E'_{ka} \qquad (37)$$

and, considering the relation (31):

$$S_g = \frac{E_{ka} + E'_{ka}}{2T_0} \tag{38}$$

with $T_0$ environmental temperature. In this case the environmental energy can be related to an energy reference state, for example the temperature of the atomic nucleus. The result obtained agrees with the previous obtained by using a quantum approach. Now, we show an example on a molecular heat engine, in order to confirm numerically our results.

**Application to molecular heat engine**

A heat engine is a machine converting heat energy into work and they are cyclic engines. First law of the thermodynamics, which is about the conversion of energy, says that difference of energy inlet to the system and energy outlet from the system for heat producing engines is equal to energy change in the system. It is expressed by the differential form of the equation (23);

$$\delta Q - \delta W = dU \tag{39}$$

For a cyclic system at the steady state conditions. Fist law of the thermodynamics is written as follows:

$$\delta Q = \delta W \tag{40}$$

For quantum thermodynamic system heat and work changes and they result [61]:

$$\delta Q = \omega dS \tag{41}$$

$$\delta W = S d\omega \tag{42}$$

where, $\omega$ is the energy-level gap and $S$ is the spin. Expectation value of spin operator is [61] :

$$S = -\frac{1}{2} Tanh\left(\frac{\beta\omega}{2}\right) \tag{43}$$

The temperature $\beta$ (throughout this paper "temperature" will refer top rather than $T$ for simplicity), if not stated otherwise ($\beta = 1/k_B T$, where $T$ is the absolute temperature) [61]. In calculations, $k_B$ and $h$ are assumed as 1 for simplicity.

Real engines operating with thermodynamic cycle always include irreversibilities based on entropy generation and lost work resulted from irreversibilities. Defining the lost work can be accomplished by comparing the heat engine with totally reversible (or reversible) cycle called as Carnot engine. Carnot heat engine is not only external reversible but internal irreversible. Another theoretical cycle is called as Curzon-Ahlborn engine that is internal reversible (endoreversible) but external irreversible. One can determined external irreversibilities by removing the work produced by endoreversible cycle from the work produced by reversible cycle. Reversible (Carnot) cycle, endoreversible and irreversible (actual heat engine) cycles (actual heat engine) for a quantum heat engine are shown in Figure 1. Using equations heat transfer equations [59], work output of reversible cycle of spin system can be written as follows:

$$Q_H^{rev} = \frac{\omega_1}{2} Tanh \frac{\beta_H \omega_1}{2} - \frac{\omega_2}{2} Tanh \frac{\beta_H \omega_2}{2} + \frac{1}{\beta_H} \ln \left( \frac{Cosh \frac{\beta_H \omega_2}{2}}{Cosh \frac{\beta_H \omega_1}{2}} \right) \quad (44)$$

$$Q_L^{rev} = \frac{\omega_4}{2} Tanh \frac{\beta_L \omega_4}{2} - \frac{\omega_3}{2} Tanh \frac{\beta_L \omega_3}{2} - \frac{1}{\beta_L} \ln \left( \frac{Cosh \frac{\beta_L \omega_4}{2}}{Cosh \frac{\beta_L \omega_3}{2}} \right) \quad (45)$$

$$W^{rev} = Q_H^{rev} - Q_L^{rev} \quad (46)$$

Similar to reversible cycle, work generated by endoreversible cycle can be obtained by as follows

$$Q_H^{endo} = \frac{\omega_1'}{2} Tanh \frac{\beta_h \omega_1'}{2} - \frac{\omega_2'}{2} Tanh \frac{\beta_h \omega_2'}{2} + \frac{1}{\beta_h} \ln \left( \frac{Cosh \frac{\beta_h \omega_2'}{2}}{Cosh \frac{\beta_h \omega_1'}{2}} \right) \tag{47}$$

$$Q_L^{endo} = \frac{\omega_4'}{2} Tanh \frac{\beta_l \omega_1'}{2} - \frac{\omega_3'}{2} Tanh \frac{\beta_l \omega_2'}{2} - \frac{1}{\beta_l} \ln \left( \frac{Cosh \frac{\beta_l \omega_4'}{2}}{Cosh \frac{\beta_l \omega_3'}{2}} \right) \tag{48}$$

$$W^{endo} = Q_H^{endo} - Q_L^{endo} \tag{49}$$

where $\omega_1', \omega_2', \omega_3', \omega_4'$ are defined in the following relation [61]:

$$\omega_1' = \frac{-2 Tanh^{-1}(2S_1)}{\beta_h}, \quad \omega_2' = \frac{-2 Tanh^{-1}(2S_2)}{\beta_h}, \quad \omega_3' = \frac{-2 Tanh^{-1}(2S_2)}{\beta_l}, \quad \omega_4' = \frac{-2 Tanh^{-1}(2S_1)}{\beta_l}$$
(35)

We investigate a two-level system and the working medium is non-interacting spin -1/2 systems. Considered heat engine cycle involve two isothermal branches connected by two irreversible adiabatic branches. Heat engine operates between heat reservoirs at $\beta_H$ and $\beta_L$, which are thermal phonon systems. The reservoirs are infinitely large and their internal relaxations are very strong. In addition, time dependent external magnetic field is applied to the system. Actual (irreversible) work output can be calculated by using the following equations:

$$Q_H^{irrev} = \frac{\omega_1'}{2} Tanh \frac{\beta_h \omega_1'}{2} - \frac{\omega_2'}{2} Tanh \frac{\beta_h \omega_2'}{2} + \frac{1}{\beta_h} \ln \left( \frac{Cosh \frac{\beta_h \omega_2'}{2}}{Cosh \frac{\beta_h \omega_1'}{2}} \right) \tag{50}$$

$$Q_L^{irrev} = \frac{\omega_4^*}{2} Tanh \frac{\beta_l \omega_4^*}{2} - \frac{\omega_3^*}{2} Tanh \frac{\beta_l \omega_3^*}{2} - \frac{1}{\beta_l} \ln \left( \frac{Cosh \frac{\beta_l \omega_4^*}{2}}{Cosh \frac{\beta_l \omega_3^*}{2}} \right) \tag{51}$$

$$W^{irrev} = Q_H^{irrev} - Q_L^{irrev} \tag{52}$$

According to the quantum adiabatic theorem, rapid change in the external magnetic field causes a quantum non-adiabatic phenomenon. The effect of the quantum non-adiabatic phenomenon on the performance characteristics of the heat engine cycle is similar to internally dissipative friction in the classical analysis. $x$ is a parameter resulted from the internal irreversibility in adiabatic branch 2-3 and 4-1. Since change of the $S_2$ to $S_3$ and $S_4$ to $S_1$ is linear, $x$ is a first grade parameter. According to this definition, $S_3$ and $S_4$ can be written as follows [59, 60]:

$$S_2 = S_3 - x \tag{53}$$

$$-\frac{1}{2}Tanh\left(\frac{\beta_H \omega_2}{2}\right) = -\frac{1}{2}Tanh\left(\frac{\beta_h \omega_3^*}{2}\right) - x \tag{54}$$

$$S_4 = S_1 - x \tag{55}$$

$$-\frac{1}{2}Tanh\left(\frac{\beta_L \omega_1}{2}\right) - x = -\frac{1}{2}Tanh\left(\frac{\beta_l \omega_4^*}{2}\right) \tag{56}$$

Lost work resulted from external irreversibility is:

$$W^{ext} = W^{rev} - W^{end} \tag{57}$$

Lost work because of external and internal irreversibilities (total irreversibilities) is calculated by removing work output of the actual heat engine from the work output obtained reversible heat engine:

$$W^{tot} = W^{rev} - W^{irrev} \tag{58}$$

Finally internal irreversibilities are calculated by removing the last work based on the external from irreversibilities the lost work based on total irreversibility:

$$W^{int} = W^{tot} - W^{ext} \tag{59}$$

According to Reference [76] entropy generation can be expressed as:

$$\frac{W^{tot}}{T_o} = \frac{E_{ka} + E_{ka}^{'}}{2T_o} \tag{60}$$

$$W^{tot} = \frac{E_{ka} + E_{ka}^{'}}{2} \tag{61}$$

where, $E_{ka}$ is the kinetic energy acquired by the atom as a consequence of the energy and momentum conservations and $E_{ka}^{'}$ is the kinetic energy of the atom in the final state, corresponding to the initial ground state of the electron involved in the transition.

External, internal and total irreversibilities are calculated respectively by using the previous relations here obtained:

$$W^{ext} = \frac{1}{2\beta_h \beta_H \beta_l \beta_L} \left\{ \begin{array}{l} -2\beta_H \beta_l \beta_L \ln\left(\frac{\sqrt{\frac{1}{1+Cosh(\beta_H \omega_1)}}}{\sqrt{\frac{1}{1+Cosh(\beta_H \omega_2)}}}\right) - 2\beta_h \beta_H \beta_L \ln\left(\frac{\sqrt{\frac{1}{1+Cosh(\beta_H \omega_2)}}}{\sqrt{\frac{1}{1+Cosh(\beta_H \omega_1)}}}\right) + 2\beta_h \beta_l \beta_L \ln\left(\frac{Cosh(\beta_H \omega_2)}{Cosh(\beta_H \omega_1)}\right) \\ +2\beta_h \beta_H \beta_l \ln\left(\frac{Cosh\left(\frac{\beta_H \beta_l \omega_1}{2\beta_L}\right)}{Cosh\left(\frac{\beta_H \beta_l \omega_2}{2\beta_L}\right)}\right) + \beta_h \beta_H \beta_l \beta_L \omega_1 Tanh\left(\frac{\beta_H \omega_1}{2}\right) + 2\beta_H(\beta_h - \beta_l)\beta_L Arctanh\left(Tanh\left(\frac{\beta_H \omega_1}{2}\right)\right) Tanh\left(\frac{\beta_H \omega_1}{2}\right) \\ -\beta_h \beta_H^2 \beta_l \omega_1 Tanh\left(\frac{\beta_H \beta_l \omega_1}{2\beta_L}\right) - \beta_h \beta_H \beta_l \beta_L \omega_2 Tanh\left(\frac{\beta_H \omega_2}{2}\right) - 2\beta_H(\beta_h - \beta_l)\beta_L Arctanh\left(Tanh\left(\frac{\beta_H \omega_2}{2}\right)\right) Tanh\left(\frac{\beta_H \omega_2}{2}\right) \\ +\beta_h \beta_H^2 \beta_l \omega_2 Tanh\left(\frac{\beta_H \beta_l \omega_2}{2\beta_L}\right) \end{array} \right\}$$

(62)

$$W^{int} = \frac{1}{\beta_l} \left\{ \begin{array}{l} \ln\left(\frac{\sqrt{\frac{1}{1+Cosh(\beta_H \omega_2)}}}{\sqrt{\frac{1}{1+Cosh(\beta_H \omega_1)}}}\right) - \ln\left(\frac{\sqrt{1-\left(Tanh\left(\frac{\beta_H \omega_2}{2}\right)-2x\right)^2}}{\sqrt{1-\left(Tanh\left(\frac{\beta_H \omega_1}{2}\right)+2x\right)^2}}\right) - Arctanh\left(Tanh\left(\frac{\beta_H \omega_1}{2}\right)\right) Tanh\left(\frac{\beta_H \omega_1}{2}\right) \\ +ArcTanh\left(2x + Tanh\left(\frac{\beta_H \omega_1}{2}\right)\right)\left(2x + Tanh\left(\frac{\beta_H \omega_1}{2}\right)\right) + ArcTanh\left(Tanh\left(\frac{\beta_H \omega_2}{2}\right)\right) Tanh\left(\frac{\beta_H \omega_2}{2}\right) \\ +ArcTanh\left(2x - Tanh\left(\frac{\beta_H \omega_2}{2}\right)\right)\left(Tanh\left(\frac{\beta_H \omega_2}{2}\right)-2x\right) \end{array} \right\}$$

(63)

$$W^{tot} = \frac{1}{2}\begin{pmatrix} \dfrac{2\ln\left(\dfrac{Cosh\left(\dfrac{\beta_H \omega_2}{2}\right)}{Cosh\left(\dfrac{\beta_H \omega_1}{2}\right)}\right)}{\beta_H} - \dfrac{2\ln\left(\dfrac{\sqrt{\dfrac{1}{1+Cosh(\beta_H \omega_1)}}}{\sqrt{\dfrac{1}{1+Cosh(\beta_H \omega_2)}}}\right)}{\beta_h} + \dfrac{2\ln\left(\dfrac{Cosh\left(\dfrac{\beta_H \beta_l \omega_1}{2\beta_L}\right)}{Cosh\left(\dfrac{\beta_H \beta_l \omega_2}{2\beta_L}\right)}\right)}{\beta_L} - \dfrac{2\ln\left(\dfrac{\sqrt{1-\left(Tanh\left(\dfrac{\beta_H \omega_2}{2}\right)-2x\right)^2}}{\sqrt{1-\left(Tanh\left(\dfrac{\beta_H \omega_1}{2}\right)+2x\right)^2}}\right)}{\beta_l} \\ +\omega_1 Tanh\left(\dfrac{\beta_H \omega_1}{2}\right) - \dfrac{2Arctanh\left(Tanh\left(\dfrac{\beta_H \omega_1}{2}\right)\right)Tanh\left(\dfrac{\beta_H \omega_1}{2}\right)}{\beta_h} + \dfrac{2Arctanh\left(2x+Tanh\left(\dfrac{\beta_H \omega_1}{2}\right)\right)\left(2x+Tanh\left(\dfrac{\beta_H \omega_1}{2}\right)\right)}{\beta_l} \\ -\dfrac{\beta_H \omega_1 Tanh\left(\dfrac{\beta_H \beta_l \omega_1}{2\beta_L}\right)}{\beta_L} - \omega_2 Tanh\left(\dfrac{\beta_H \omega_2}{2}\right) + \dfrac{2Arctanh\left(Tanh\left(\dfrac{\beta_H \omega_2}{2}\right)\right)Tanh\left(\dfrac{\beta_H \omega_2}{2}\right)}{\beta_h} \\ +\dfrac{2Arctanh\left(2x-Tanh\left(\dfrac{\beta_H \omega_2}{2}\right)\right)\left(Tanh\left(\dfrac{\beta_H \omega_2}{2}\right)-2x\right)}{\beta_l} + \dfrac{\beta_H \omega_2 Tanh\left(\dfrac{\beta_H \beta_l \omega_2}{2\beta_L}\right)}{\beta_L} \end{pmatrix}$$

(64)

In Table 1 the parameters used in calculations are listed. After calculations work outputs for reversible ($W^{rev}$), endoreversible ($W^{endo}$) and irreversible cycles ($W^{irrev}$) are found and shown in Table 2. Calculated results for the quantum heat engine are matching with results of a macro scale heat engine; reversible work has the biggest, endoreversible is the second and work output of the irreversible cycle is the smallest one. This proves that quantum thermodynamics and classical thermodynamics are accordance with each other. Using work outputs of reversible, endoreversible and irreversible cycles, total, external and internal irreversibilities are determined. Results show that external irreversibilities of the quantum cycle are much more than (more than three times) internal irreversibilities. It can be said that irreversibilities resulted from the finite temperature heat transfer more effective than irreversibilities based on quantum friction. In addition to that, we define total irreversibilities in terms of kinetic energies described in [75], which is footprint of the process.

**4. Conclusions**

Famous Carnot engine is a totally reversible, which represents upper limits for heat engines. This means it has maximum efficiency and maximum work output. However, Carnot heat engine is theoretical and ,unfortunately, it is impossible that heat engines operates without losses in reality. In other words, there are always losses in actual thermal cycles and reason of these losses are resulted from the entropy generation. These losses, which cause to decrease in work output and efficiency, are called as lost work, irreversibilities or exergy destruction. These irreversibilities may be divided into internally and externally based. As it referred before, Carnot heat engine is totally reversible, there is no irreversibility or lost work in it, endoreversible heat engine is described as internally reversible (however, externally irreversible). In classical thermodynamics, relationship between these heat engines can be defined and values of irreversibilities might be calculated.

In this study, lost works are determined for molecular heat engine that is -1/2 spin system and it is seen that results correspond to macro heat engines. In addition, our results are similar with ref [98-102], which are about miro/molecular heat engines and quantum thermodynamics. It is recommended that nano/quantum thermodynamics should be focused on because of advances in nano technolgy.

The relation obtained show that during any process, there always exist a type of irreversibility, due to interaction of the matter with the electromagnetic waves of the environment, even if frictions don't exist. This irreversibility is related to the nature itself of the matter, and it is due to the existence of the spontaneous flows between open systems and environment due to thermal disequilibrium. Consequently, part of the energy absorbed by the system is converted in atomic irreversibility and cannot be used to convert the absorbed heat into work. Here we have suggested a new approach to explain the macroscopic irreversibility in thermodynamics by introducing an energy footprint in quantum mechanics and deriving a change in the Schrödinger's equation for a Hydrogen-like atom. We have followed the Einstein's, Schrödinger's and Gibbs considerations on the interaction between particles and thermal radiation (photons), which leads to consider the atom as an open system in interaction with an external flows of photons. In conclusion, we state that the quantum mechanical analysis shows that particle path information isn't preserved because the particle interactions with photons in the thermal radiation field change the internal states of the particles themselves with a microscopically irreversibility, in accordance with the irreversible measurements that John von Neumann showed increase the entropy [103]. The results here obtained is a confirmation of the hypothesis that an energy footprint due to irreversibility exists also in atomic level transition [104-108]. Irreversibility is the result of the interaction between open system and its environment.

**Table 1. Parameters used in calculations**

| Parameter | Unit | Value |
|-----------|------|-------|
| $\beta_H$ | $J^{-1}$ | 0.0100 |
| $\beta_L$ | $J^{-1}$ | 0.0333 |
| $\beta_h$ | $J^{-1}$ | 0.0111 |
| $\beta_l$ | $J^{-1}$ | 0.0313 |
| $x$ | - | 0.0003 |
| $\omega_1$ | J | 5 |
| $\omega_2$ | J | 3 |

**Table 2. Calculated values for the cycle**

| Parameter | Unit | Value |
|-----------|------|-------|
| $W^{rev}$ | J | 0.0141 |
| $W^{endo}$ | J | 0.0116 |
| $W^{irrev}$ | J | 0.0108 |
| $W^{ext}$ | J | 0.0025 |
| $W^{int}$ | J | 0.0008 |
| $W^{tot}$ | J | 0.0033 |

**Figure 1. Carnot quantum heat engine, endoreversible quantum heat engine and irreversible quantum heat engine**

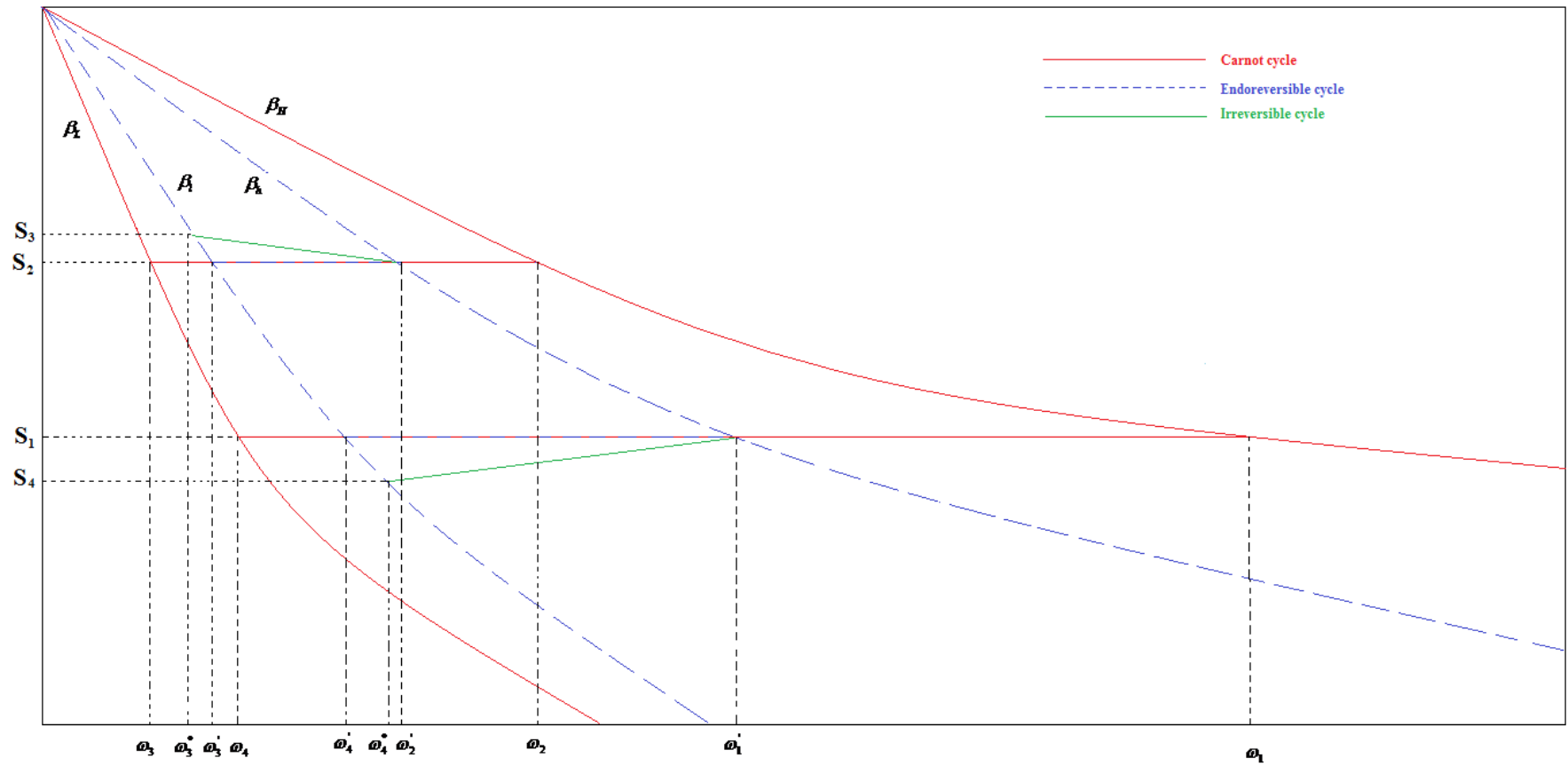